\documentclass[a4paper,11pt]{article}
\usepackage{pos}

\title{The QCD topological susceptibility at high temperatures via staggered fermions spectral projectors}
\ShortTitle{$\chi_{\mathrm{QCD}}$ at high $T$ via staggered fermions spectral projectors}

\author[a,b]{Andreas Athenodorou}
\author*[c]{Claudio Bonanno}
\author[a]{Claudio Bonati}
\author[d]{Giuseppe Clemente}
\author[a]{Francesco D'Angelo}
\author[a]{Massimo D'Elia}
\author[a]{Lorenzo Maio}
\author[e]{Guido Martinelli}
\author[f]{Francesco Sanfilippo}
\author[g,h,i]{Antonino Todaro}

\affiliation[a]{Università di Pisa and INFN Sezione di Pisa, Largo B.~Pontecorvo 3, I-56127 Pisa, Italy}
\affiliation[b]{Computation-based Science and Technology Research Center, The Cyprus Institute, 20 Kavafi Str., Nicosia 2121, Cyprus}
\affiliation[c]{INFN Sezione di Firenze, Via G.~Sansone 1, Sesto Fiorentino, I-50019 Firenze, Italy}
\affiliation[d]{Deutsches Elektronen-Synchrotron (DESY), Platanenallee 6, 15738 Zeuthen, Germany}
\affiliation[e]{Dipartimento di Fisica and INFN Sezione di Roma ``La Sapienza'', Piazzale Aldo Moro 5, I-00185 Rome, Italy}
\affiliation[f]{INFN Sezione di Roma Tre, Via della Vasca Navale 84, I-00146 Rome, Italy}
\affiliation[g]{Department of Physics, University of Cyprus, P.O.~Box 20537, 1678 Nicosia, Cyprus}
\affiliation[h]{Faculty of Mathematics und Natural Sciences, University of Wuppertal, Wuppertal-42119, Germany}
\affiliation[i]{Dipartimento di Fisica, Università di Roma ``Tor Vergata'', Via della Ricerca Scientifica 1, I-00133 Rome, Italy}
\affiliation[]{}

\emailAdd{a.athenodorou@cyi.ac.cy}
\emailAdd{claudio.bonanno@fi.infn.it}
\emailAdd{claudio.bonati@unipi.it}
\emailAdd{giuseppe.clemente@desy.de}
\emailAdd{francesco.dangelo@phd.unipi.it}
\emailAdd{massimo.delia@unipi.it}
\emailAdd{lorenzo.maio@phd.unipi.it}
\emailAdd{guido.martinelli@roma1.infn.it}
\emailAdd{francesco.sanfilippo@infn.it}
\emailAdd{atodar01@ucy.ac.cy}

\abstract{The QCD topological observables are essential inputs to obtain theoretical predictions about axion phenomenology, which are of utmost importance for current and future experimental searches for this particle. Among them, we focus on the topological susceptibility, related to the axion mass. We present lattice results for the topological susceptibility in QCD at high temperatures obtained by discretizing this observable via spectral projectors on eigenmodes of the staggered Dirac operator, and we compare them with those obtained with the standard gluonic definition. The adoption of the spectral discretization is motivated by the large lattice artifacts affecting the standard gluonic susceptibility, related to the choice of non-chiral fermions in the lattice action.}

\FullConference{41$^{\text{st}}$ International Conference on High Energy physics, ICHEP2022\\ 6$^{\text{th}}$-13$^{\text{th}}$ July, 2022\\Bologna, Italy}

\usepackage{braket}
\usepackage{placeins}
\usepackage{slashed}

\newcommand{\beq}{\begin{eqnarray}}
\newcommand{\eeq}{\end{eqnarray}}
\newcommand{\beqnn}{\begin{eqnarray*}}
\newcommand{\eeqnn}{\end{eqnarray*}}
\newcommand{\Tr}{\ensuremath{\mathrm{Tr}}}
\newcommand{\round}{\ensuremath{\mathrm{round}}}
\newcommand{\stag}{\ensuremath{\mathrm{stag}}}
\newcommand{\SP}{\ensuremath{\mathrm{SP}}}
\newcommand{\cool}{\ensuremath{\mathrm{cool}}}
\newcommand{\gluo}{\ensuremath{\mathrm{gluo}}}
\newcommand{\DIGA}{\ensuremath{\mathrm{DIGA}}}
\newcommand{\clov}{\ensuremath{\mathrm{clov}}}
\newcommand{\CP}{\ensuremath{\mathrm{CP}}}

\begin{document}
\maketitle

\section{Introduction}\label{sec:intro}

General theoretical arguments predict that QCD allows for violations of the $\CP$ symmetry via the dimensionless parameter $\theta$ coupling the integer-valued topological charge
\beq\label{eq:topocharge_continuum}
Q=\frac{1}{32 \pi^2} \varepsilon_{\mu\nu\rho\sigma}\int \Tr\{G^{\mu\nu}(x)G^{\rho \sigma}(x)\} d^4x \in \mathbb{Z},
\eeq
to the QCD action, with $G^{\mu\nu}$ the gluon field-strength tensor. So far, no experimental evidence of strong $\CP$ violation has ever been found, suggesting that $\theta$ is actually vanishing; in particular, the most precise experimental upper bound on the neutron electric dipole moment reflects in the stringent upper bound $\vert \theta \vert \lesssim O(10^{-9}, 10^{-10})$. This fine-tuning issue is known as \emph{strong $\CP$ problem} and is currently one of the most intriguing questions left open by the Standard Model.

Among the many Beyond Standard Model solutions that have been considered in the literature, the \emph{Peccei--Quinn axion}~\cite{Peccei:1977hh} is one of the most promising. This hypothetical particle dynamically relaxes $\theta$ to zero thanks to the properties of the \emph{Peccei--Quinn symmetry}, and is at the same time a possible Dark Matter candidate.

An interesting theoretical feature of axion models is that the knowledge of the QCD \emph{topological susceptibility} $\chi\equiv\braket{Q^2}/V$, with $V$ the space-time volume, allows to put upper bounds on the scale $f_a$ at which the Peccei--Quinn symmetry spontaneously breaks, via the relation $m_a^2 f_a^2 = \chi$, with $m_a$ the effective axion mass. This is a piece of information of the utmost importance, being it a necessary input for current and future experimental searches for axions. More precisely, for the purpose of axion cosmology, what is needed is the behavior of $\chi$ for high values of the temperature $T$ (corresponding to early times of the Universe evolution).

In recent years, Monte Carlo simulations on the lattice have been extensively employed to compute $\chi(T)$, being topological properties purely non-perturbative features of gauge theories. The lattice computation of the topological susceptibility in the deconfined phase poses however several non-trivial numerical challenges. In this paper we are concerned in particular with the problem of suppressing the large lattice artifacts that affect the computation of $\chi$ at finite lattice spacing $a$ via standard methods~\cite{Bonati:2015vqz,Borsanyi:2016ksw,Petreczky:2016vrs}. Achieving such a reduction is useful to perform more reliable extrapolations of this quantity towards the continuum limit $a \to 0$, especially at high $T$.

In this respect, a recent proposal involves the adoption of spectral projectors~\cite{Giusti:2008vb, Luscher:2010ik, Cichy:2015jra, Alexandrou:2017bzk} over the eigenvectors of the Dirac operator $\slashed{D}$ to define the topological charge. Such definition relies on the \emph{index theorem}, which relates $Q$ to the spectral properties of the low-lying eigenvalues of $\slashed{D}$, and has proven to be effective in reducing the magnitude of finite-lattice-spacing corrections to the continuum limit of $\chi$ at zero temperature~\cite{Alexandrou:2017bzk}.

This paper reports on the main results of Ref.~\cite{Athenodorou:2022aay}. In this work, spectral projectors in the presence of staggered quarks~\cite{Bonanno:2019xhg} are employed to provide controlled continuum extrapolations of $\chi$ in high-$T$ QCD at the physical point for a range of temperatures going from $\sim200$ to $\sim600$~MeV. The main results of~\cite{Athenodorou:2022aay} have also been presented during this year's \emph{LATTICE} conference~\cite{pos:lattice22_spectral}.

This manuscript is organized as follows: in Sec.~\ref{sec:num_setup} we summarize the main aspects regarding the staggered fermion spectral projectors approach; in Sec.~\ref{sec:results} we present our results for the QCD topological susceptibility at high temperature; finally, in Sec.~\ref{sec:conclusions} we draw our conclusions.

\section{Numerical setup}\label{sec:num_setup}

We discretize QCD at finite temperature $T=(a N_t)^{-1}$ on a $N_t \times N_s^3$ lattice with lattice spacing $a$ using the tree level Symanzik-improved gauge action for the gluon sector and $2+1$ flavors of rooted stout staggered fermions for the quark one. The continuum limit is approached along a Line of Constant Physics (LCP), meaning that the bare coupling $\beta$ and the bare quark masses $m_l,m_s$ are tuned to ensure that simulations at each lattice spacing are carried on at the physical point, i.e., for physical pion mass $m_\pi \simeq 135$~MeV and for physical strange-to-light quark mass ratio $m_s/m_l \simeq 28.15$.

Topological charge~\eqref{eq:topocharge_continuum} is defined on the lattice via staggered spectral projectors. According to the index theorem, only zero-modes (i.e., with non-vanishing chiralities) contribute to $Q$. Staggered quarks however explicit break the chiral symmetry for non-vanishing $a$, and no zero-mode is present in the spectrum of $D_\stag$. Thus, the sum over the chiralities of zero-modes has to be extended at finite lattice spacing as~\cite{Bonanno:2019xhg}:
\beq
Q = \sum_{\lambda=0} u_\lambda^\dagger \gamma_5 u_\lambda \quad \longrightarrow \quad Q_{\SP,0} = \frac{1}{n_t} \sum_{\vert \lambda \vert \le M} u_\lambda^\dagger \gamma_5^{(\stag)} u_\lambda,
\eeq
where $Q_{\SP,0}$ is the bare spectral projectors charge, $\gamma_5^{(\stag)}$ is the staggered version of the Dirac matrix $\gamma_5$, the factor $1/n_t=1/4$ cancels out staggered taste degeneration. The eigenvectors $u_\lambda$ satisfy
\beq
i D_\stag u_\lambda = \lambda u_\lambda, \qquad \lambda \in \mathbb{R},
\eeq
with $D_\stag$ matching the same discretization of the lattice action, mentioned at the beginning of this section. The bare spectral projectors charge is renormalized as~\cite{Bonanno:2019xhg}:
\beq
Q_{\SP} = Z_Q^{(\stag)} Q_{\SP,0}, \qquad \quad Z_Q^{(\stag)} = \sqrt{\frac{\braket{\Tr\left\{\mathbb{P}_M\right\}}}{\braket{\Tr\left\{\gamma_5^{(\stag)} \mathbb{P}_M \gamma_5^{(\stag)} \mathbb{P}_M\right\}}}},
\eeq
where $\mathbb{P}_M \equiv \sum_{\vert \lambda \vert \le M} u_\lambda u_\lambda^\dagger$ is the staggered spectral projector over eigenvectors with eigenvalues lying below $M$. The topological susceptibility is finally given by $\chi_{\SP} \equiv \braket{Q_{\SP}^2}/{V}$, where $V=a^4 N_t N_s^3$ is the lattice volume in physical units.

The threshold mass $M$ cutting-off spectral sums is a free parameter of this definition, as its particular value becomes irrelevant in the continuum limit, where chiral symmetry is restored and only zero-modes contribute to $Q$. However, a prescription to keep its renormalized value in physical units constant as the continuum limit is approached is needed to guarantee $O(a^2)$ corrections to the continuum value $\chi$:
\beq
\chi_{\SP}(a,M) = \chi + c(M_R) a^2 + o(a^2).
\eeq
Staggered quarks allow to adopt this simple prescription: since $M$ renormalizes as a quark mass $m_f$, and since the continuum limit is approached along a LCP, it is sufficient to keep $M/m_f = M_R/m_f^{(R)}$ constant along the LCP for any given flavor $f$ to ensure that the continuum limit is approached at constant $M_R$. In the following we will express $M$ in terms of the strange quark mass $m_s$.

\section{Results}\label{sec:results}

\begin{figure}[!htb]
\centering
\includegraphics[scale=0.38]{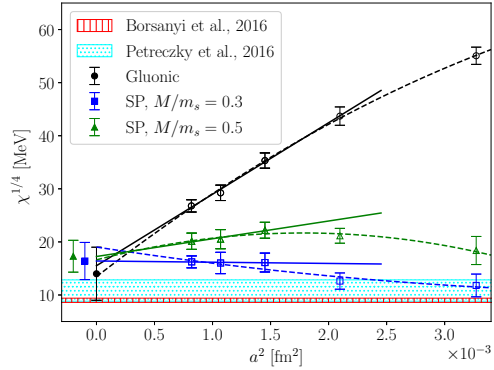}
\caption{Figure taken from Ref.~\cite{Athenodorou:2022aay}. Comparison of continuum extrapolations of $\chi^{1/4}_{\SP}$ determinations for two choices of $M/m_s$ (squares and triangles) with the extrapolation of $\chi_\gluo^{1/4}$ (circles) and the results for $\chi^{1/4}$ of Refs.~\cite{Borsanyi:2016ksw, Petreczky:2016vrs} (dotted and dashed shaded areas). Figure refers to $T\simeq 430$~MeV.}
\label{fig:continuum_limit_t_430}
\end{figure}

In Fig.~\ref{fig:continuum_limit_t_430} we show an example of continuum extrapolation of $\chi_{\SP}$ for a temperature $T\simeq 430$~MeV. Our spectral results for $2$ choices of $M/m_s$ are compared with the gluonic determination $\chi_{\gluo}=\braket{Q_{\gluo}^2}/V$, where $Q_{\gluo}$ is obtained computing the \emph{clover} discretization of Eq.~\eqref{eq:topocharge_continuum}
\beq
Q_\clov = \frac{-1}{2^9 \pi^2}\sum_{x}\sum_{\mu\nu\rho\sigma=\pm1}^{\pm4}\varepsilon_{\mu\nu\rho\sigma}\Tr\left\{\Pi_{\mu\nu}(x)\Pi_{\rho\sigma}(x)\right\},
\eeq
with $\Pi_{\mu\nu}(x)$ is the plaquettes built on the $\mu-\nu$ plane and rooted in the lattice site $x$, on cooled configurations after $n_\cool=80$ steps, and rounding it to the nearest integer as:
\beq
Q_\gluo = \round\left\{\alpha Q_\clov^{(\cool)}\right\}, \qquad \quad \alpha = \min_{x>1} \left\langle\left(x Q_\clov^{(\cool)} - \round\left\{x Q_\clov^{(\cool)}\right\}\right)^2\right\rangle.
\eeq
In Fig.~\ref{fig:continuum_limit_t_430} we also report the results of Refs.~\cite{Borsanyi:2016ksw, Petreczky:2016vrs}. We observe that it is possible to reduce lattice artifacts with a suitable choice of $M$, and that continuum extrapolations of $\chi_{\SP}$ for different $M/m_s$ agree within the errors. Spectral determinations are also compatible with $\chi_{\gluo}$ and with the results of Refs.~\cite{Borsanyi:2016ksw, Petreczky:2016vrs}.

\begin{figure}[!htb]
\centering
\includegraphics[scale=0.38]{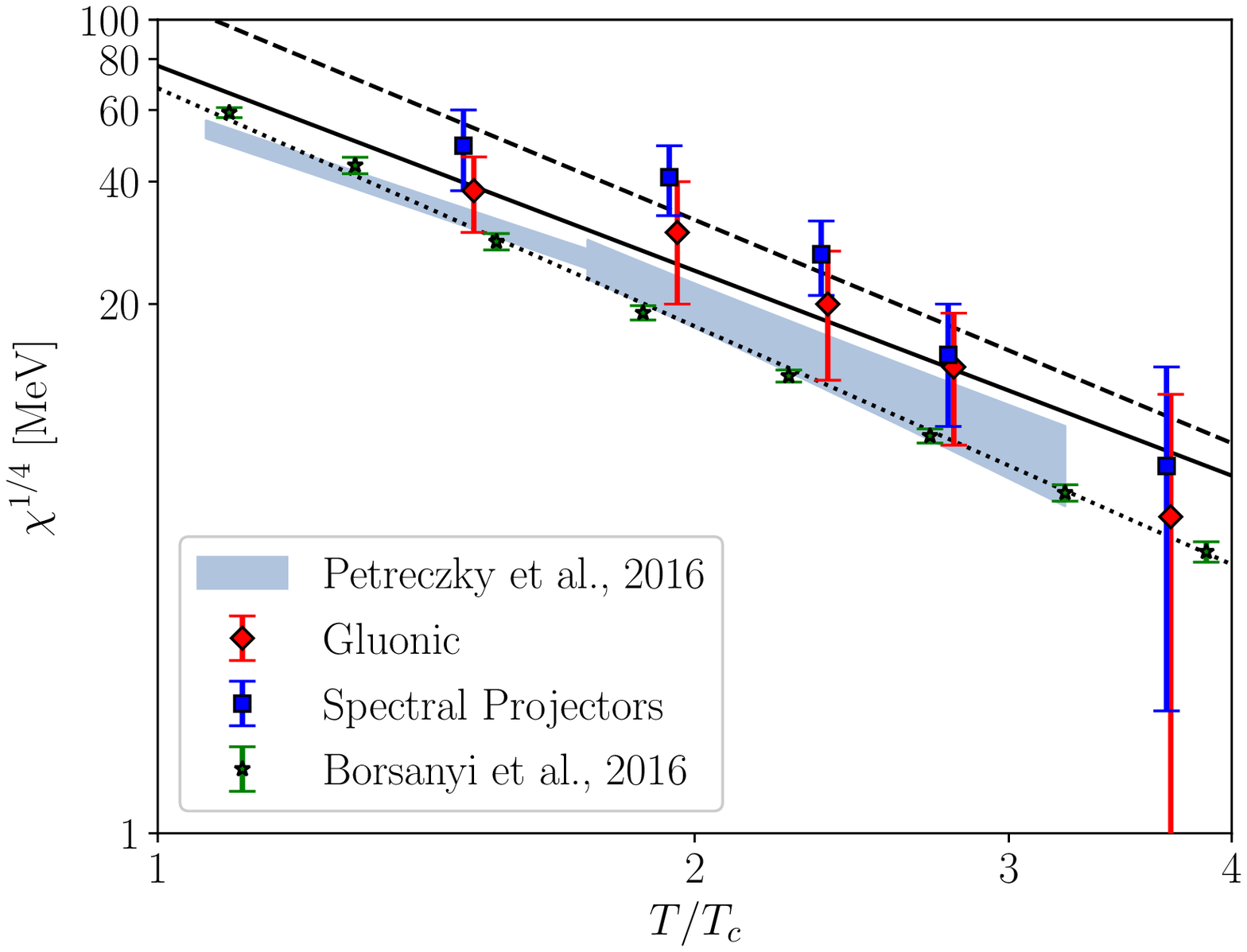}
\includegraphics[scale=0.38]{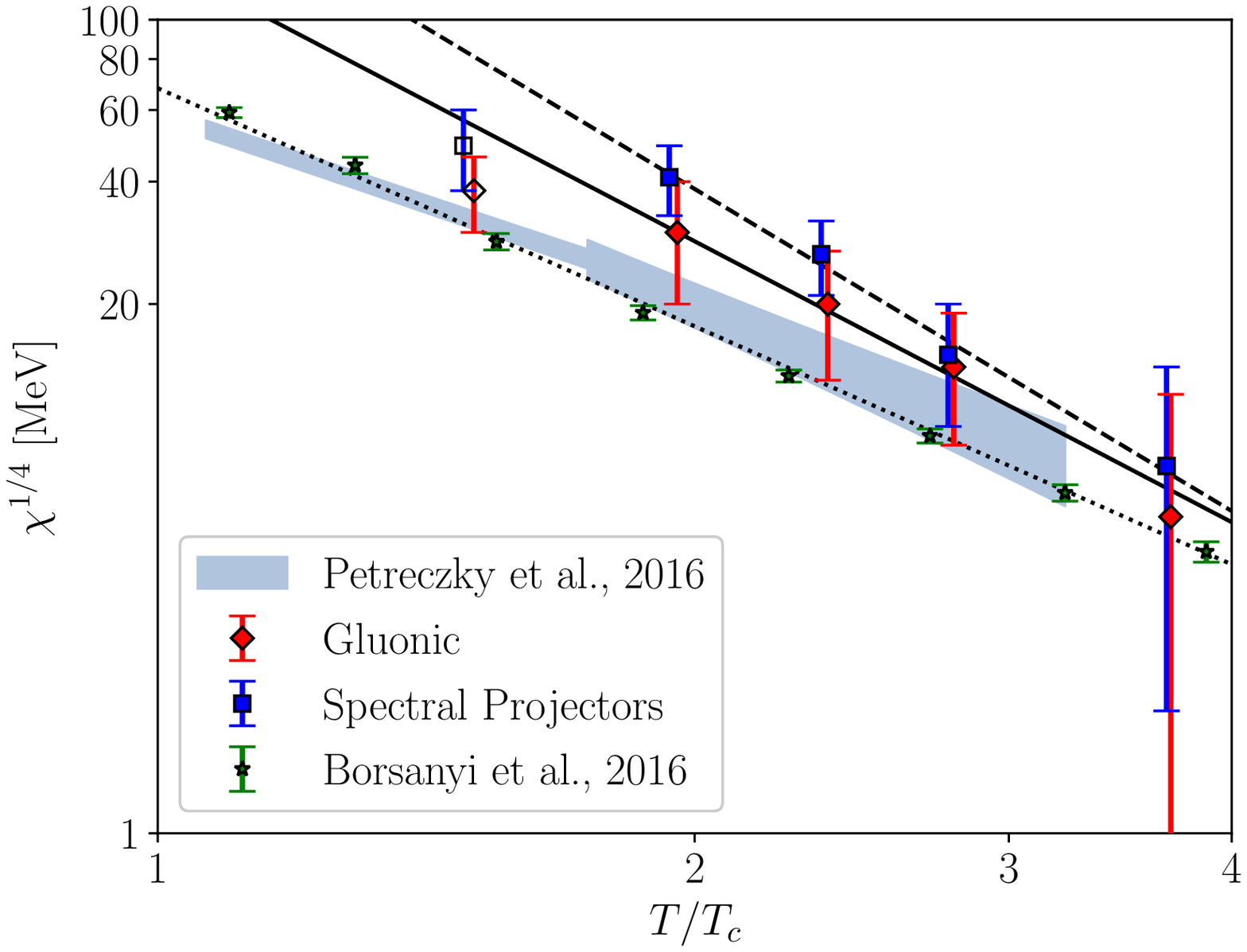}
\caption{Figures taken from Ref.~\cite{Athenodorou:2022aay}. Behavior of $\chi_{\SP}^{1/4}$ (squares) and $\chi_{\gluo}^{1/4}$ (diamonds) as a function of $T/T_c$ in $\log$-$\log$ scale, compared with results of~\cite{Borsanyi:2016ksw} (stars) and of~\cite{Petreczky:2016vrs} (shaded area). Dashed, solid and dotted lines represent best fits of $\chi_\SP$, $\chi_\gluo$ and Ref.~\cite{Borsanyi:2016ksw} data assuming the DIGA-inspired ansatz $\chi^{1/4} = A(T/T_c)^{-b}$.}
\label{fig:chi_vs_T}
\end{figure}

Our continuum extrapolations of $\chi_{\SP}$ and $\chi_\gluo$ for all the explored temperatures are shown in Fig.~\ref{fig:chi_vs_T}, along with results of Refs.~\cite{Borsanyi:2016ksw, Petreczky:2016vrs}. Error bars of $\chi_\SP$ data here reported are estimated keeping into account any systematic variation observed in the continuum extrapolation when changing $M/m_s$ within a reasonable range (see Ref.~\cite{Athenodorou:2022aay} for an extensive discussion on this point).

Both our spectral and gluonic data can be well described with a decaying power law $\chi^{1/4} \sim (T/T_c)^{-b}$, with $T_c \simeq 155$~MeV the crossover temperature, in agreement with the Dilute Instanton Gas Approximation (DIGA) prediction~\cite{Gross:1980br}, cf.~Fig.~\ref{fig:chi_vs_T}. Exponents also agree very well with the prediction $b_{\DIGA}=2$:
\begin{align*}
b_\SP = 1.82(43)&, \quad b_\gluo = 1.67(51),& \quad\quad &\text{fit for}~T\ge230~\mathrm{MeV};\\
b_\SP = 2.63(81)&, \quad b_\gluo = 2.3(1.1),& \quad\quad &\text{fit for}~T\ge300~\mathrm{MeV}.
\end{align*}
However, we observe that the DIGA-like power law seems to set in for higher values of $T$ compared to findings of Ref.~\cite{Borsanyi:2016ksw}. Moreover, we observe that our results lie systematically above those of Refs.~\cite{Borsanyi:2016ksw, Petreczky:2016vrs}, and in particular we observe respectively $\sim3$ and $\sim 2.5$ standard deviation tensions among our and Refs.~\cite{Borsanyi:2016ksw, Petreczky:2016vrs}'s determinations in the range $300\lesssim T \lesssim 365$~MeV, see Fig.~\ref{fig:chi_vs_T}.

\FloatBarrier

\vspace*{-1.14\baselineskip}
\section{Conclusions}\label{sec:conclusions}

This paper reports on the main findings of~\cite{Athenodorou:2022aay}. The spectral projectors definition of the topological susceptibility allows to improve the convergence of this quantity towards the continuum limit, as the large lattice artifacts affecting the standard gluonic definition can be reduced by a suitable choice of the threshold mass $M$ used to cut-off the spectral sums defining $\chi_\SP$.

Our continuum-extrapolated spectral determination of $\chi(T)$ is well described by a decaying power-law $\chi^{1/4}(T) \sim (T/T_c)^{-b}$ as predicted by the DIGA, with $b$ also agreeing with the prediction $b_\DIGA=2$. We find a $\sim 2-3$ standard deviation tension with results of Refs.~\cite{Borsanyi:2016ksw, Petreczky:2016vrs} in the range $300~\text{MeV}\lesssim T \lesssim 365$~MeV, which deserves to be better clarified with future dedicated studies.

To this end, it would be interesting to refine our determinations for $T \lesssim 400$~MeV by adding finer lattice spacings to our analyses to improve our determinations of the continuum limit of $\chi_{\SP}$. It would also be interesting to push the present comparison towards higher temperatures, in particular around the $T\sim 1$~GeV scale, which is also relevant for axion cosmology.

To reach higher temperatures and finer lattice spacings, however, the infamous \emph{topological freezing} problem has to be faced, as it is well known that the Critical Slowing Down experienced by standard lattice Monte Carlo algorithms close to the continuum limit is particularly severe for topological quantities. A promising candidate to face this problem is the \emph{parallel tempering on boundary conditions} algorithm proposed by M.~Hasenbusch for $2d$ $\CP^{N-1}$ models~\cite{Hasenbusch:2017unr} and recently adopted both in the latter case~\cite{Berni:2019bch} and in purely-gluonic $4d$ Yang--Mills theories~\cite{Bonanno:2020hht, Bonanno:2022yjr} to efficiently mitigate the topological freezing issue.

\providecommand{\href}[2]{#2}\begingroup\raggedright\endgroup

\end{document}